\documentclass[doublecol]{epl2} 
\usepackage[main=english,vietnamese]{babel}
\usepackage{lipsum}
\usepackage{curves}
\usepackage{times}
\usepackage{graphicx,rotating}
\usepackage{xcolor}
\usepackage[version=4]{mhchem}

\usepackage{graphicx, amsmath, mhchem}
\usepackage{subfigure, amssymb}
\usepackage{float} % Put the figures where you want them to be! {figure}[H]
\usepackage[makeroom]{cancel}
\usepackage{color}

\graphicspath{ {plots/} }

\newcommand{\I}{\mathrm{i}}
\newcommand{\D}{\mathrm{d}}

\DeclareMathAlphabet{\vecfont}{OT1}{cmr}{bx}{it}
\renewcommand{\vec}[1]{\vecfont{#1}}
  
\title{The Debye layer as a transmission line in the 4 Hz - 100 kHz frequency range}
\author{Thanh-Trị Châu\inst{1} \and Giovanni Zocchi\inst{1} }
\shortauthor{T. T. Châu \etal}

\institute{                    
	\inst{1} Department of Physics and Astronomy, University of California, Los Angeles  - California 90095, USA
}

\abstract{
We report measurements on the dynamics of the Debye layer at a gold electrode in several electrolytes. In the experiments, 
the Debye layer transmits a damped voltage wave along the electrode, which we use to probe the dynamics. We compare 
the measurements with traditional impedance models, which schematize the Debye layer as a capacitance. We find good agreement 
for very dilute electrolytes, but also for an ionic liquid. However, the same model fails for the concentrated electrolyte, as 
ion - solvent - ion interactions become important. }

\begin{document}

\maketitle
\section{Introduction}
\begin{figure}[h]
	\onefigure[width=1\linewidth]{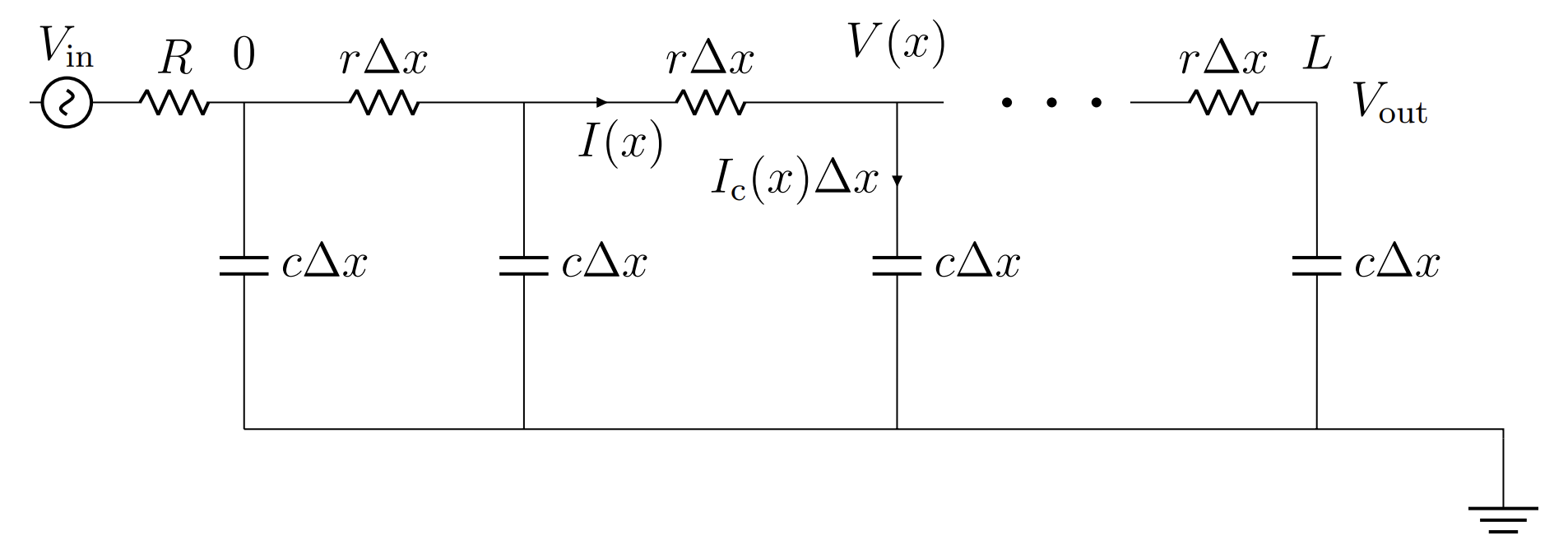}
	\caption{Schematic of the Debye layer at the gold electrode as a distributed RC transmission line in 1D.  
	The resistive path in the gold is of resistance $r$ per unit length and the electrolyte within the Debye layer is considered as a capacitor 
	with capacitance $c$ per unit length. $R$ is a trailing resistance, present in the experiments, between the input AC source 
	and the starting point $x=0$ of the line. In the experiments, we measure the amplitude and phase of 
	$V_{\mathrm{out}}$ relative to $V_\mathrm{in}$, at $x = L$. }
	\label{fig.1.}
\end{figure}

A charged surface in an electrolyte drives the formation of a cloud of counterions within a thin boundary layer (Debye layer) next to the surface. 
This electric double layer (EDL) has a profound effect on macromolecular and colloidal interactions. Ionic screening 
transforms the long range Coulomb interaction into a short range force. The large electric field within the Debye layer (typically of order 
$100 \, \mathrm{mV} / 1 \, \mathrm{nm}$) can affect a variety of chemical and physical processes. For example, the binding dynamics 
of biological macromolecules involves the disruption of the EDL at the surface of contact. Action potential transmission 
is accompanied by a large perturbation of the EDL at the cell membrane \cite{Bazant2020}. Another motivation 
for the study of EDL dynamics lies in the development of capacitive energy storage devices (supercapacitors) 
and electrochemical processes for energy conversion \cite{Pilon2015, Wu2022}. \\ 
Direct experimental measurements of the Debye layer are challenging, due to the $\mathrm{nm}$ scale. 
The static properties, such as the concentration profile of ions in the diffuse layer near a charged surface, 
can be deduced from measurements of the force between such surfaces \cite{Israelachvili_Book, Gebbie2017}. The corresponding theoretical framework 
is based on the Poisson-Boltzmann equation and known as the Gouy-Chapman theory. 
Measurements of the dynamics are mostly obtained from electrochemical cells. There are three common methods \cite{Pilon2015}. 
In Electrochemical Impedance Spectroscopy (EIS) a voltage consisting 
of a DC component plus a sine wave is imposed; the measured quantity is the current. The measurements are often presented in terms of 
a complex impedance vs frequency (a Nyquist plot). Cyclic Voltammetry (CV) is essentially the same measurement with a different voltage protocol 
(a triangular wave). In Galvanometric Cycling the electrode current is imposed and the measured quantity is the voltage. Physical quantites 
of interest, such as the EDL capacitance, are extracted by comparing to a model \cite{Pilon2015, Wu2022}. 
Increasingly, models are compared with MD simulations, which provide quantities (such as the structure of the EDL) only indirectly 
accessible to experiments \cite{Noh2019, Park2022}. \\
To a first approximation, and for non-Fradaic processes, the EDL behaves like a capacitor, charging and discharging 
as charges on the two sides of the electrode-electrolyte interface accumulate and disperse without an actual charge transfer across the interface. 
To introduce the characteristic scales, let us consider a planar electrode in a 1:1 electrolyte (such as \ce{NaCl} in water). 
From the linearized 
Debye-Falkenhagen equation for the charge density $\rho$:
\begin{equation}
		\frac{\partial \rho}{\partial t} - \chi k_\mathrm{B}T \, \nabla^2 \rho + \frac{2\chi}{\epsilon} |e|^2 n_0  \, \rho = 0
		\label{eq: DF1} 
\end{equation}
where $\chi$ is the mobility of the ions (for simplicity assumed the same for the two species e.g. \ce{Na+} and \ce{Cl-}), $n_0$ the 
bulk concentration of \ce{NaCl}, $|e|$ the proton charge and $\epsilon$ the dielectric constant of the medium, one obtains   
the characteristic length scale $\ell = \sqrt{(\epsilon k_\mathrm{B} T) / (2 |e|^2 n_0)}$ (the Debye length) and time scale $\tau = \ell^2/(\chi k_\mathrm{B}T)=\epsilon / (2 |e|^2 n_0 \chi) = \epsilon / \sigma$ ($\sigma$ is the ionic conductivity of the solution). Below we write the equations in dimensionless form using these characteristic scales and scaling $\rho$ by $2|e|n_0$. 
In an axis-symmetric geometry with the $z$ axis perpendicular to the electrode, 
(\ref{eq: DF1}) becomes: 
\begin{equation}
	\frac{\partial \rho}{\partial t} - \frac{\partial^2 \rho}{\partial z^2} + \rho = 0\,.
	\label{eq: DF2}
\end{equation}
The electrostatic potential $\phi$ is related to $\rho$ through the Poisson equation; 
scaling $\phi$ by $k_\mathrm{B}T/|e|$, it reads: 
\begin{equation}
	\frac{\partial^2 \phi}{\partial z^2}=-\rho\,.
	\label{eq: Poisson}
\end{equation}
With sinusoidal forcing $\phi (z = 0, t) = \exp (\I\omega t)$ and $\phi,\, \rho \rightarrow 0$ for $z \rightarrow \infty$ (in the bulk), the solution of (\ref{eq: DF2}) and (\ref{eq: Poisson}) 
is: 
\begin{eqnarray}
	\phi(z,t)&=&  e^{- k z} e^{\I\omega t}\,,\label{eq: DFphisol}\\\
	\rho (z, t) &=& -k^2e^{- k z} e^{\I\omega t}\,,\label{eq: DFrhosol}\\
	\text{with}\quad &&k^2 = 1 + \I \omega\,.\nonumber
\end{eqnarray} 
We are concerned only with the regime $\omega \ll 1$ , in which case $k \approx 1 + \I\omega / 2$ and (\ref{eq: DFrhosol}) describes 
the formation of a Debye layer of size one (size $\ell$ in dimensional variables), independent of frequency (apart from a slight phase lag). 
The reason is that, taking as an example a $150 \, \mathrm{mM}$ (``physiological'') \ce{NaCl} solution, $\sigma \approx 10 \, \mathrm{mS/cm}$, $n_0 \approx 10^{20} \, \mathrm{cm}^{-3}$ so that $\tau \approx 1 \, \mathrm{ns} $ and $\ell \approx 1 \, \mathrm{nm} $ . \\
%\red{Given that $\tau$ is measured in nanoseconds, we deviate only slightly from the equilibrium described by the Poisson-Boltzmann equation, which is the static case of the Debye-Falkenhagen equation.}\\
%\red{Given that $\tau$ is measured in nanoseconds, we effectively sample a huge order of microstates within a period of the driving voltage of the frequencies of interest here. The mean-field Debye-Falkenhagen equation is justified as a first approximation.} \\ 
The capacitive current at the electrode is obtained by considering the relation of the surface charge to the electric field: 
$\epsilon E_n = Q/A $ where $Q / A$ is the charge per unit area and $E_n$ the normal component of the electric field 
at the surface. The capacitive current 
density $j$ is then $j = (\partial / \partial t) (Q / A) = \epsilon (\partial / \partial t) E_n = 
- \epsilon  (\partial / \partial t) (\vec{\nabla} \phi \cdot \vec{n})_{z=0} $ 
with $\vec{n}$ the unit normal. Using (\ref{eq: DFphisol}) in the regime $\omega \ll 1$ we then find 
$|j| \approx \epsilon\,\omega (1 + \omega^2 / 8)$, to be compared with a driven RC circuit, where, in the same approximation, 
the current is $|I| \approx C \omega (1 - \omega^2 / 2)$. We see that the frequency behavior of the capacitive current at the electrode 
departs from that of an RC circuit only for $\omega \sim 1$. For this reason one associates to the Debye layer a capacitance per unit area 
$c = |j| / |\dot{\phi}(z=0)| \approx \epsilon (1 + \omega^2 / 8)$ which, for $\omega \ll 1$, is the constant $\epsilon$ 
(or $\epsilon / l$ in dimensional variables). 
However, this simple theory neglects hydrodynamic effects, as well as interactions between ions, and between ions and the electrode. \\ 
Here we probe Debye layer dynamics through an experimental configuration where the Debye layer forms part of an RC transmission line. 
We measure voltage transmission at the end of the line, given a sinusoidal input at the other end. 
The Debye layer at a planar gold electrode provides the capacitive part of the line, while the resistive part is provided by the thin film 
gold layer which forms the electrode. It is helpful to reason in terms of the continuum limit of the equivalent discrete elements circuit. 
As a first approximation, Fig.~\ref{fig.1.} is a schematic of the (1D) transmission line if the EDL can be considered as a distributed capacitance  
with capacitance per unit length $c$\,; $r$ is the resistance per unit length of the gold electrode. 
The potential along the line satisfies the diffusion equation
\begin{equation}
	\frac{\partial V(x,t)}{\partial t} - \frac{1}{rc} \frac{\partial^2 V(x,t)}{\partial x^2} = 0\,.
\end{equation}

\noindent For AC driving voltage $V(x=0, t) = V_{\mathrm{in}} e^{\I \omega t}$ with the Ansatz
\begin{equation}
	V(x,t)=\exp(-kx)\exp(\I\omega t) 
\end{equation} 

\noindent one obtains a complex wave vector
\begin{equation}\label{eq.2}
	k= (1+\I)\sqrt{\frac{\omega rc}{2}}\,.
\end{equation}
In this simple model, for increasing frequency the output voltage $V_\mathrm{out}$ has an amplitude going exponentially to zero 
and a monotonously decreasing phase (modulo $2\pi$).

\section{Materials, sample preparation and experimental procedure}
\begin{figure}[h]
	\onefigure[width=0.9\linewidth]{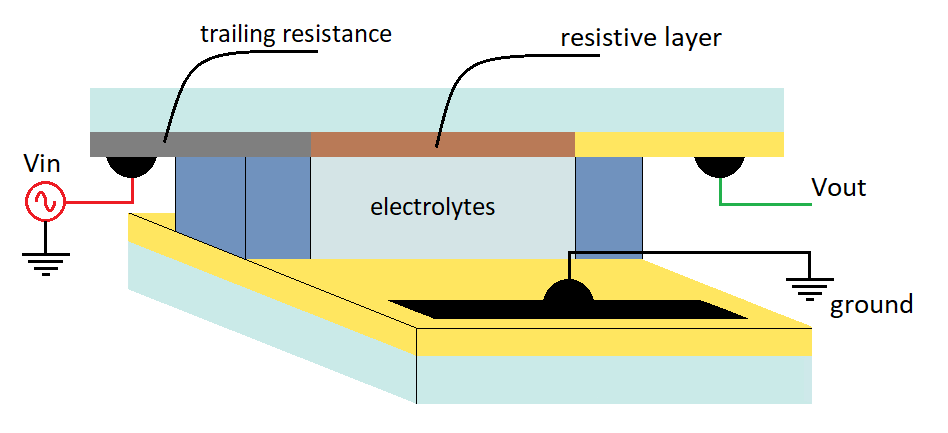}
	\caption{Schematic of an experimental sample. The electrolyte is sealed in a $120 \, \mathrm{\mu m}$ thick cell obtained from two microscope 
	slides separated by spacers. The inner surface of the slides is conductive through a thin layer of gold evaporated on it. One electrode 
	serves as the ground, while the other provides a resistive path for the transmission line.}
	\label{fig.2}
\end{figure}

The electrolyte solutions under study are \ce{NaCl} aqueous solutions of $10 \, \mathrm{mM}$, $100 \, \mathrm{mM}$ and $1 \, \mathrm{M}$ concentrations in 
$10 \, \mathrm{mM}$ \ce{NaH2PO4}/\ce{Na2HPO4}  buffer with pH $7$ at $ 25 $ \textcelsius. We choose phosphate buffer because its $\mathrm{pK_a}$ at the second step is close to the neutral $\mathrm{pH}$ and has modest variation with temperature.\footnote{$\mathrm{pK_a2=7.21}$ with a variation of approximately $-0.0028/\text{\textcelsius}$.}  The ionic liquid is 1-Butyl-3-methylimidazolium hexafluorophosphate (\ce{C8H15F6N2P}). All chemicals are from Sigma-Aldrich. \\
The ionic liquid or salt solution is sealed with a UV-activated epoxy (Loctite AA 3525) between two gold-coated glass slides arranged 
perpendicular to one another and separated by two spacer strips $120 \, \mathrm{\mu m}$ thick (see Fig.~\ref{fig.2}). 
The top plate provides a resistive path for the voltage signal and is prepared by depositing $3 \, \mathrm{nm}$ of \ce{Cr} followed by 
$20 \, \mathrm{nm}$ of \ce{Au} on a glass slide using electron-beam evaporation method (CHA Mark 40). The bottom plate, having an extra 
$10 \, \mathrm{nm}$ of \ce{Au} compared to the top, serves as the ground electrode. The contacts are made by soldering gauge-28 wires 
on a buffering layer of electrically conductive silver paint (MG Chemicals 842AR-P) applied on the gold layer to prevent the latter from being peeled off. \\
We use a lock-in amplifier (SR850) for the measurements. The reference signal from the lock-in drives  an in-house built voltage clamp circuit 
which applies the input voltage between $V_\mathrm{in}$ and ground, as in Fig.~\ref{fig.2}. By switching the input to the detection path of the lock-in between $V_\mathrm{in}$ and $V_\mathrm{out}$, we measure both their amplitude and phase relative to the reference. We deduce from these measurements their relative amplitude and phase difference. For each experimental condition, we keep the temperature of the sample 
stable within $\pm 0.01 \text{\textcelsius}$ in an in-house built thermoelectrically controlled chamber.

\section{Results}
\begin{figure}
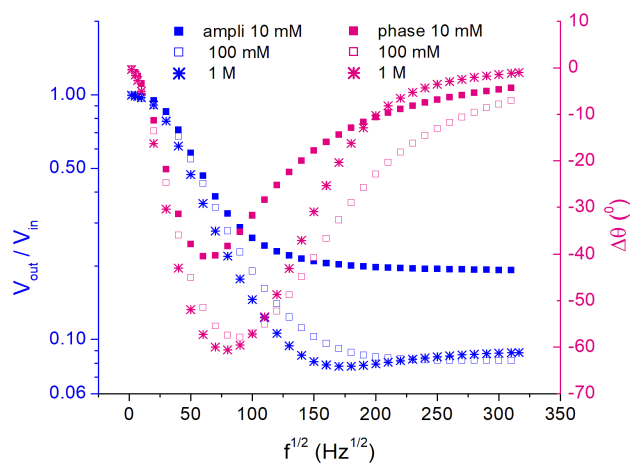

	\onefigure[width=\linewidth]{plots/Fig_3_P.png}
	\caption{Measured amplitude (blue) and phase (red) of $V_{\mathrm{out}} / V_{\mathrm{in}}$ vs the square root of the driving frequency $f = \omega / 2 \pi$ 
	for three concentrations of buffered \ce{NaCl} solution: $10 \, \mathrm{mM}$ (filled squares), $100 \, \mathrm{mM}$ (empty squares), and $1 \, \mathrm{M}$ 
	(stars). The amplitude is shown on a log scale, the phase on a linear scale. The driving amplitude was $V_{\mathrm{in}} = 24 \, \mathrm{mV}\,\mathrm{rms}$, 
	and the temperature $25 $ \textcelsius .}
	\label{fig.3}
\end{figure}

\begin{figure}[h]
	\onefigure[width=0.9\linewidth]{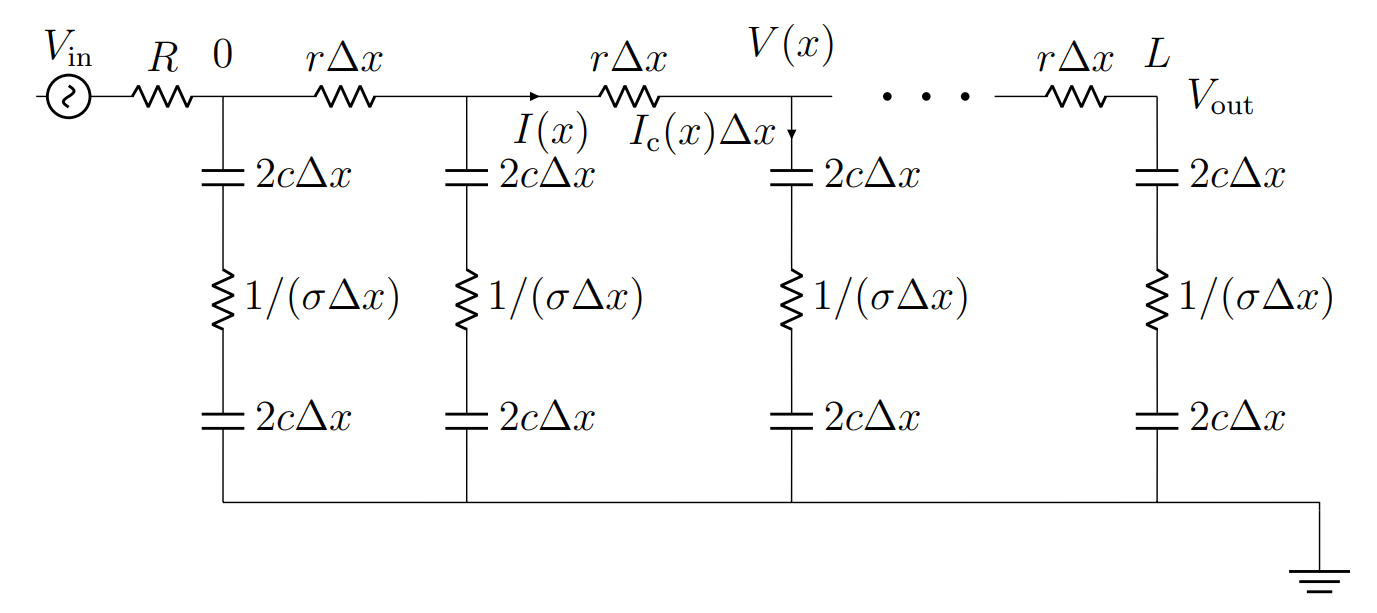}
	\caption{Schematic of the 1D transmission line model where the electrolyte is considered as two capacitors sandwiching a series resistor. 
	Each capacitor has a value of $2c$ per unit length, whereas the resistor corresponds to a conductance of $\sigma$ per unit length. 
	The transmission line electrode has resistance per unit length $r$, whereas the ground electrode is an equipotential, due to the thicker 
	gold layer on it. }
	\label{fig.4}
\end{figure}

Fig.~\ref{fig.3} shows the ratio of the rms amplitudes of the voltage at the end of the electrolytic cell, $V_\mathrm{out}$, to the input voltage 
$V_\mathrm{in}$, as a function of the square root of the driving frequency (blue symbols). Also shown is the phase of $V_\mathrm{out}$ relative to 
$V_\mathrm{in}$ (red symbols). Measurements are displayed for 3 different \ce{NaCl} concentrations of the electrolyte: 10, 100, and 1000 $\mathrm{mM}$. 
In the text and in the figures we refer to the above \ce{NaCl} concentrations to identify the samples, however all these solutions also contain 
$10 \, \mathrm{mM}$ phosphate buffer, so for example the total ionic strength of the ``$10 \, \mathrm{mM}$ salt'' samples is actually $20 \, \mathrm{mM}$. 
We notice immediately that at high enough frequencies the behavior of both the amplitude and phase of the output signal is 
non monotonic with salt concentration; this feature cannot be explained by a model based on the mean field theory of the Introduction.  
Equation (\ref{eq.2}) predicts that the amplitudes in Fig.~\ref{fig.3} should decrease linearly on this log-linear scale above some small frequency\footnote{The behavior at low frequencies reflects the finite size of the transmission line, among other things.}.
Similarly the phase should decrease linearly on the same scale with jumps from $-\pi$ to $\pi$ due to its periodic nature. Up to some moderate frequency, such as $110^2= 12.1 \, \mathrm{kHz}$ for $100 \, \mathrm{mM}$, this is indeed the case. However at higher frequency, the amplitude saturates to some constant level 
while the phase goes back up to zero. The reason is that in the experiment there are {\it two} Debye layers, one at the ``live'' and one at the 
ground electrode, connected by a resistive path through the electrolyte. 
A more appropriate transmission line model is therefore as shown in Fig.~\ref{fig.4}, where the ground electrode is endowed with its own 
(capacitive) Debye layer and the conductance (per unit length) $\sigma$ refers to conduction through the electrolyte in the direction orthogonal 
to the plates. 

\noindent We now consider this 1D transmission line, of finite length $L$, and solve analytically  for the output voltage $V_\mathrm{out}\equiv V(L)$,  
under harmonic driving. Ohm's law relates the current $I(x)$ through the resistive gold layer to the voltage $V(x)$ at an arbitrary point 
along the transmission line as
\begin{equation}
	I(x)=-\frac{1}{r}\frac{\partial V(x)}{\partial x}\,.
\end{equation}
We atttribute the spatial variation of this current to a capacitive current per unit length $I_\mathrm{c}(x)$, related to the local voltage $V(x)$ 
through the impedance of the electrolyte strip of width $\D x$: 
\begin{equation}
	-\frac{\partial I(x)}{\partial x}=I_\mathrm{c}(x)=V(x)\left(\frac{1}{\I\omega c}+\frac{1}{\sigma}\right)^{-1}\,.
\end{equation}
As a result, the voltage $V(x)$ satisfies
\begin{equation}
\frac{\partial^2 V(x)}{\partial x^2}=\frac{\I\omega rc\sigma}{\I\omega c+\sigma}V(x) 
\label{eq:transmission_line}
\end{equation}
where $\omega$ is the forcing frequency. 
We solve eq. (\ref{eq:transmission_line}) under boundary conditions that account for a vanishing current at $x=L$ 
\begin{equation*}
	V'(L)\sim I(L)=0
\end{equation*}
and a voltage drop across the trailing resistance $R$
\begin{equation*}
	V_\mathrm{in}+\frac{R}{r}V'(0)=V(0)\,.
\end{equation*} 
The result for the complex output $V_\mathrm{out}$ is 
\begin{equation}\label{eq.3}
\frac{V_\mathrm{out}}{V_\mathrm{in}}=\frac{1}{\cosh(kL)+\alpha kL\sinh(kL)},
\end{equation}
with 
\begin{equation}
	\alpha=\frac{R}{rL}
\label{eq:alpha}
\end{equation} 
being the ratio between the trailing resistance $R$ and the resistance of the metal layer in direct contact with the electrolyte. \\
The real and imaginary parts of the wave vector $k$ are: 
\begin{equation}
\begin{split}
	k'L & =\left[\frac{\omega/\omega_{rc}\left(\sqrt{1+\omega^2/\omega_{c\sigma}^2}+\omega/\omega_{c\sigma}\right)}{2\left(1+\omega^2/\omega_{c\sigma}^2\right)}\right]^{1/2}\\
		k''L & =\left[\frac{\omega/\omega_{rc}\left(\sqrt{1+\omega^2/\omega_{c\sigma}^2}-\omega/\omega_{c\sigma}\right)}{2\left(1+\omega^2/\omega_{c\sigma}^2\right)}\right]^{1/2}\,.
\end{split}
\label{eq:kr_ki}
\end{equation}

\noindent The frequencies $\omega_{rc}$ and $\omega_{c\sigma}$ are set by the charging time of the capacitors, and limited by the 
(longitudinal) resistance of the gold layer and the (transverse) resistance of the electrolyte, respectively: 
\begin{eqnarray}
	\label{eq:omega_rc}
		\omega_{rc}&=&\frac{1}{(rL)(cL)}\\
		\omega_{c\sigma}&=&\frac{\sigma}{c} \, .
	\label{eq:omega_cs}
\end{eqnarray}
$\alpha$ is not a fitting parameter as we can deduce it from measurements of the resistive gold layer's geometry and surface resistance. For a rectangular gold layer $2.98\pm 0.30 \,\mathrm{cm}$ long and $0.93\pm 0.10\,\mathrm{cm}$ wide, we measure a resistance of $9.3\pm 0.1\,\mathrm{\Omega}$. This corresponds to a surface resistance of $2.9\pm0.4\,\mathrm{\Omega}/\mathrm{sq}$ ({$\mathrm{sq}$ refers to any square patch of surface). We are left with two fitting parameters: $\omega_{rc}$ and $\omega_{c\sigma}$.

\begin{figure}[h]
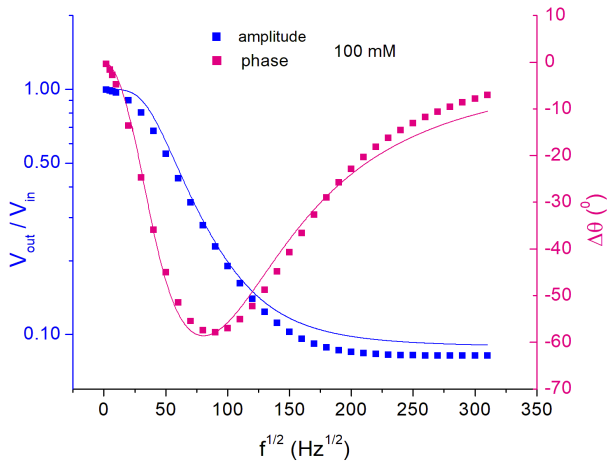

	\onefigure[width=\linewidth]{plots/Fig_5_P.png}
	\caption{Amplitude and phase of $V_{\mathrm{out}} / V_{\mathrm{in}}$ for sample A: 
	$100 \, \mathrm{mM}$ \ce{NaCl} in $10 \, \mathrm{mM}$ phosphate buffer, pH 7 at $25$ \textcelsius \, (same data as in Fig. \ref{fig.3}). The frequency sweep 
	is carried out at $25$ \textcelsius \, and $24 \, \mathrm{mV}$ rms driving voltage. The solid lines show the global two-parameter fit 
	according to Eq.~(\ref{eq.3}), using a measured $\alpha=1.69$.}
	\label{NaCl_100_fit}
\end{figure}

\begin{figure}[h]
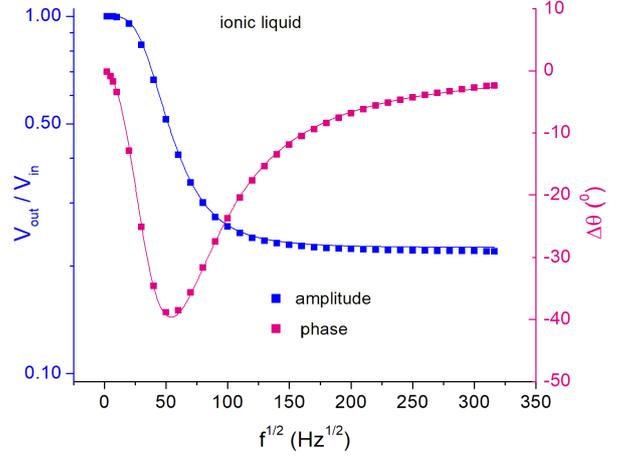

	\onefigure[width=\linewidth]{plots/Fig_6_P.png}
	\caption{Amplitude (blue) and phase (red) of $V_{\mathrm{out}} / V_{\mathrm{in}}$ for the ionic liquid \ce{C8H15F6N2P}, at $25$ \textcelsius \, and 
	$24 \, \mathrm{mV}$ rms driving voltage. Solid lines show the two parameter fit to the model Eq.~(\ref{eq.3}), using a measured $\alpha=1.29$.}
	\label{Ionic_liquid_fit}
\end{figure} 

\noindent The two parameters fit to the model of Eq. (\ref{eq.3}) is displayed in Fig. \ref{NaCl_100_fit} 
for the $100 \, \mathrm{mM}$ salt solution (same data as in Fig. \ref{fig.3}). The fit for the $10 \, \mathrm{mM}$ salt is noticeably better than the case shown, 
while for the $1 \, \mathrm{M}$ case it is noticeably worse (data not shown). 
The conclusion we draw is that the model of Fig. \ref{fig.4}, which schematizes the Debye layer as a fixed 
(frequency independent) capacitance and the bulk electrolyte as a resistance, describes the frequency dependence fairly well. 
However we also see systematic departures from the data, at the level of $10 \, \%$ for the $100 \, \mathrm{mM}$ salt 
and $20 \, \%$ for the $1 \, \mathrm{M}$ concentration. The discrepancy between model and experiments is significant 
as the stability and reproducibility of the experiment (for the same sample) 
is of order the size of the dots in the plots. It cannot be attributed to nonlinearities arising from voltage dependent 
processes (such as chemical reactions): the measurements of Fig. \ref{fig.3}, obtained for $V_{\mathrm{in}} = 24 \, \mathrm{mV}$, were repeated 
(on the same samples) for $V_{\mathrm{in}} = 50 \, \mathrm{mV}$, and identical results (overlapping symbols, not shown) were obtained. 
Intriguingly, the measurements on the ionic liquid, shown in Fig. \ref{Ionic_liquid_fit}, show no discrepancy at all 
with the model (\ref{eq:transmission_line}). \\ 
However, the virtues of the RC transmission line model are somewhat tarnished if one tries to directly relate its effective parameters to the 
physical properties of the electrolytes. Table 1 summarizes the parameters obtained from the fits, for the 3 salt concentrations 
and the ionic liquid. The $100 \, \mathrm{mM}$ salt 
condition was measured for two independent samples A and B. The two fitting parameters are $\omega_{rc}$ and $\omega_{c \sigma}$ 
(Eq. (\ref{eq:kr_ki})); $\alpha$ is obtained from the measured conductivity of the gold layer and the geometry of the sample. 
The capacitance per unit area {\it of one EDL}, $C$, is obtained from the value of $\omega_{rc}$ using (\ref{eq:omega_rc}) 
and the measured geometry (length and width) of the cell. For comparison, the next column in the Table lists the corresponding 
capacitance value $\epsilon / \ell$ obtained from the (calculated) Debye length $\ell$. Note that for the $1 \, M$ salt and 
more so for the ionic liquid, the calculated $\ell$ is smaller than the size of the ions. 
The last two columns display the bulk conductivity 
of the electrolyte obtained from $ \omega_{c \sigma}$ using (\ref{eq:omega_cs}) and the cell geometry, and for comparison the 
literature values. The quantity $\epsilon / \ell$, which is proportional to $\sqrt{\epsilon}$, is calculated using $\epsilon = 80$ (relative to the permittivity of free space)
for the salt solutions and $\epsilon = 11$ for the ionic liquid \cite{Baker2001}. 
Focusing on the capacitance $C$, we see that in all cases the EDL capacitance measured in the experiment is more than an order of magnitude 
lower than $\epsilon / \ell$, which is the value calculated assuming a parallel plates capacitor of thickness 
$\ell$ (the Debye length). 
This well known phenomenon is usually attributed to the existence of a Stern layer of immobilized water molecules and counterions 
at the metal surface. In terms of the model of Fig. \ref{fig.4} the effect is to add a ``Stern layer capacitance'' 
$C_\mathrm{S} = \epsilon_\mathrm{S} / \delta_\mathrm{S}$ in series to the Debye layer capacitance $C$; $\epsilon_\mathrm{S}$ is the dielectric constant of the Stern 
layer, $\delta_\mathrm{S}$ its thickness \cite{Pilon2015}. The composite EDL capacitance is then $C_\mathrm{EDL} = (C C_\mathrm{S}) / (C + C_\mathrm{S}) < C_\mathrm{S}$ 
i.e. it is bounded by $C_\mathrm{S}$. 
With representative values $\epsilon_\mathrm{S} \approx 2$ and $\delta_\mathrm{S} \approx 4 \mathrm{\AA}$ for the Stern layer (see for example the 
detailed analysis in \cite{Park2022}) one obtains $C_\mathrm{S} \approx 4.4 \, \mathrm{\mu F / cm^2}$, not inconsistent with our measured 
values for the \ce{NaCl} electrolytes. \\ 
For the conductivity $\sigma$, at low ionic strength (the $10 \, \mathrm{mM}$ sample) there is rough agreement between the value obtained 
from the experiment (which assumes that the resistors labeled $1 / (\sigma \Delta x)$ in Fig. \ref{fig.4} correspond to the conductivity 
of the bulk electrolyte) and the actual bulk conductivity of the electrolyte. But for high ionic strength there is a discrepancy of more 
than an order of magnitude. On the other hand, the ionic liquid again seems to display ``ideal'' behavior in that there is rough agreement 
between the experimental value and the actual bulk conductivity. \\ 
Next to ionic strength, temperature is another thermodynamic control parameter in the experiments, affecting the Debye layer, the Stern 
layer, and the bulk conductivity, among other factors. Without providing a comprehensive analysis in this Letter, we show in 
Fig. \ref{temperature_variation} a series of measurements at different temperatures, for the $100 \, \mathrm{mM}$ salt. 
The data for  $25$, $15$, $5$, and $0$ $^{\circ} \mathrm{C}$ are 
from the same sample, while $- 10$ and $- 15$ $^{\circ} \mathrm{C}$ are from another. The monotonic rise of the high frequency amplitude plateau 
with decreasing temperature, and corresponding behavior of the phase, are due to the decrease in the conductivity of the electrolyte 
(due to the decreased ion mobility) with decreasing temperature. At $- 15$ $^{\circ} \mathrm{C}$ the sample is frozen; there is no Debye layer 
(the conductivity $\sigma$ and capacitance per unit length $c$ are essentially zero) and we obtain a featureless response 
($V_{\mathrm{out}} / V_{\mathrm{in}} \approx 1$ and $\Delta \theta \approx 0$). Note that under our conditions the $100 \, \mathrm{mM}$ samples freeze at about 
$- 10$ $^{\circ} \mathrm{C}$, because the cell keeps the sample at approximately constant volume and therefore under pressure when frozen. 

\begin{figure}[h]
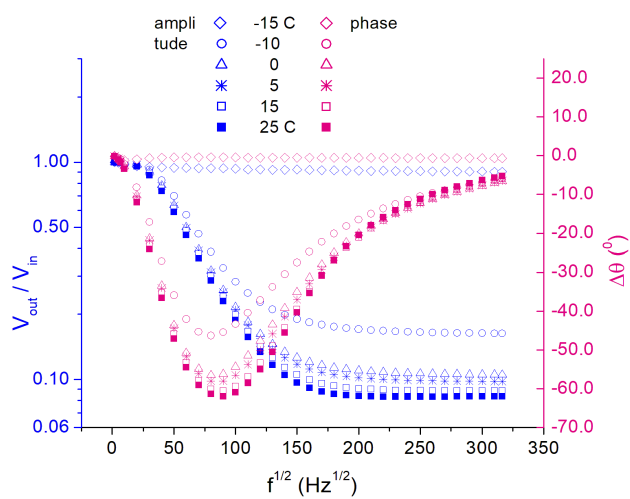

	\onefigure[width=\linewidth]{plots/Fig_7_P.png}
	\caption{Effect of temperature on the amplitude and phase of $V_{\mathrm{out}} / V_{\mathrm{in}}$ , measured for the $100 \, \mathrm{mM}$ \ce{NaCl} solution; 
	drive amplitude $V_{\mathrm{in}} = 24 \, \mathrm{mV}$. The behavior is monotonic with temperature; the different plots correspond to 
	$T = 25 ^{\circ} \mathrm{C}$ (filled squares), and $15 , 5 , 0 , -10 , -15 \, ^{\circ} \mathrm{C}$. At $ -15 \, ^{\circ} \mathrm{C}$ (open diamonds) 
	the sample is frozen (whereas at $-10 \, ^{\circ} \mathrm{C}$ (open circles) it is still liquid, due to the increased pressure). }
	\label{temperature_variation}
\end{figure}

\begin{table*}
\caption{The parameters $\omega_{rc}$ and $\omega_{c \sigma}$ (see (\ref{eq.3}), (\ref{eq:kr_ki})) 
obtained from fitting the data to the model Eq. (\ref{eq.3}) , 
for the $10 \, \mathrm{mM}$ , $100 \, \mathrm{mM}$ (two different samples), and $1 \, \mathrm{M}$  \ce{NaCl} electrolytes, and the ionic liquid. The parameter $\alpha$ is measured 
from the conductivity of the gold electrodes and the geometry of the cells (see (\ref{eq:alpha})). The capacitance per unit area $C$ is derived from 
$\omega_{rc}$ (see (\ref{eq:omega_rc})), the ionic conductivity $\sigma$ from $\omega_{c \sigma}$ (see (\ref{eq:omega_cs})). For 
comparison, the column $\epsilon / \ell$ displays the capacitance per unit area derived from the calculated Debye length $\ell$, while the 
column $\sigma_{ref}$ displays literature values of the ionic conductivity. Note that in the Table, C is the capacitance of {\it one} Debye layer, 
corresponding to $2 c \, \Delta x$ in Fig. \ref{fig.4}.}
\label{tab.1}
\begin{center}
\begin{tabular}{ |c|c|c|c|c|c|c|c| }
 \hline 
  & $\alpha$ & $\omega_{rc} \, [\mathrm{kHz}]$ & $\omega_{c \sigma} \, [\mathrm{kHz}]$ & $C$ [$\mathrm{\mu F / cm^2}$] & $\epsilon / \ell$ 
  & $\sigma \, [\mathrm{mS / cm}]$ & $\sigma_{ref}$\\ 
  \hline
 10 mM & 1.79 & 33.7 & 47.4 & 3.32 & 33 & 0.94 & 1.2\\ 
 100 mM (A) & 1.69 & 28.8 & 88.4 & 3.90 & 76 & 2.07 & 10.7\\ 
 100 mM (B) & 1.44 & 22.9 & 82.1 & 4.90 &   & 2.41 &  \\ 
 1 M & 2.11 & 22.2 & 62.5 & 5.08 & 230 & 1.90 & $\sim$ 90\\ 
 Ionic Liquid & 1.29 & 20.3 & 31.4 & 5.52 & 192  & 1.04 & 1.5\\ 
 \hline
\end{tabular}
\end{center}
\end{table*}

\section{Summary and Discussion}
We have introduced a setup in which the Debye layer at the electrode-electrolyte interface forms part of a transmission line 
along the electrode. Measurements of the voltage along the electrode, alternative to traditional current measurements 
with equipotential electrodes, reflect the dynamics 
of the Debye layer. We present measurements for buffered \ce{NaCl} solutions of different ionic strengths, in the low voltage regime where 
hydrolysis and other chemical reactions are unimportant. We also present a set of measurements for an ionic liquid.  
We find that the traditional view of the EDL as a (frequency independent) capacitance describes the dynamics fairly well in some cases, 
but not others. Specifically, for the low salt concentration electrolyte ($10 \, \mathrm{mM}$ \ce{NaCl}), and, at the other end, for the ionic liquid, 
the frequency dependence of the measurements is well described by the model of Fig. \ref{fig.4}. However, for the more concentrated 
$100 \, \mathrm{mM}$ salt electrolyte, and more so for the $1 \, \mathrm{M}$ concentration, there are large 
discrepancies between the measurements and the model. The two parameters measured from the fits, $\omega_{rc}$ and 
$\omega_{c \sigma}$, can be converted, within this model, into an effective capacitance for the EDL and an effective ionic conductivity 
for the electrolyte. In all cases, the capacitance values thus measured are an order of magnitude or more smaller than the values 
calculated for a parallel plates capacitor of thickness a Debye length $\ell$ , and using the bulk static dielectric constant of the 
medium. This behavior is attributed to the existence of a Stern layer of immobilized electrolyte molecules at the electrode 
\cite{Stern1924, Park2022}. The conductivity values measured from the fits are roughly consistent with the actual bulk ionic conductivity 
of the electrolyte for the very dilute salt solution ($10 \, \mathrm{mM}$) and for the ionic liquid. However for the more concentrated salt solutions 
the conductivity from the fits clearly does not represent the bulk conductivity. Phenomenologically one could possibly invoke 
a reduced mobility of the ions in the region of the Stern / Debye layer. The more fundamental conclusion is however 
that the impedance model Fig. \ref{fig.4} misses part of the physics. \\  
The hydrodynamics of interacting ions close to a surface forms indeed 
an interesting mathematical problem, due to the range of scales: the Debye layer at the $\mathrm{nm}$ scale, the bulk electrolyte at the $\mathrm{\mu m}$ 
or $\mathrm{mm}$ scale. The method of matched asymptotic expansions has been used in this context \cite{Bazant2004, Zhao2011}. 
The impedance elements models (such as Fig. \ref{fig.4}) circumvent the mathematical difficulties by placing a resistance 
($1/\sigma$ in Fig. \ref{fig.4}) for the bulk electrolyte, but this is unsatisfactory in general.  
The transmission line model based on impedances was  introduced in the 60's to describe electrochemistry at a porous electrode \cite{de_Levie1963}, 
a subject of renewed interest today \cite{Huang2020}. Remarkably, this simple approach seems to work well for the ionic liquid.  
Ion - solvent interactions are absent in this case, a situation analogous to the ideal behavior of a polymer melt. 
Moreover, measurements with the surface force apparatus indicate that ionic liquids behave as dilute electrolytes in terms of the static 
properties of the diffuse layer \cite{Gebbie2013, Gebbie2015}, apparently for the reason that only a small fraction of the charges 
are effectively dissociated. \\ 
From a purely experimental point of view, looking at the columns $C$ and $\sigma$ in Table 1, one would think that what is 
measured are properties of the electrode rather than the different electrolytes. In fact it is well known that even in the static case, 
the structure of the EDL is more complicated than the result of the mean field Gouy-Chapman-Stern theory suggests \cite{Wu2022}. 
For high ionic strength, the ion density profile away from the electrode is oscillatory rather than monotonic \cite{Park2022}, while the 
Debye length may increase with salt at high enough concentrations \cite{Smith2016}. The non-monotonicity with increasing salt 
concentration visible in the plots of Fig. \ref{fig.3} may be related to this latter phenomenon; further measurements with this setup 
at higher salt concentrations could be informative. Similarly it should be interesting to compare our measurements with 
the predictions from continuum theories which take into account the finite size of the ions and ion-ion interactions \cite{Bazant2011, Zhao2011}. \\

\noindent Measurements with this setup may be extended in a number of ways. A DC bias can be added to the drive, in order to probe the dynamics 
with different electrode potentials. This is routinely done in EIS, where a third (reference) electrode is normally used to standardize the 
measurements. The high voltage regime ($|e| V / k_\mathrm{B}T > 1$) will introduce nonlinearities and 
eventually new processes (water hydrolysis, potentially 
surface remodelling), and remains to be studied in this system. Coupling this transmission line configuration to redox chemical reactions in the 
electrolyte will generate reaction-diffusion systems where voltage is one phase space coordinate. These should be interesting as 
pattern forming systems. 

\begin{acknowledgments}
\noindent We thank Anastassia Alexandrova for suggesting the ionic liquid measurements. 
\end{acknowledgments}

\end{document}